\documentclass[letter]{aa}
\usepackage{graphicx}
\usepackage{txfonts}
\usepackage{epsfig}
\begin{document}
\title{AGILE detection of variable $\gamma$-ray activity from the blazar
  S5~0716+714 in September--October 2007 \thanks{The optical data presented in
  this paper are stored in the GASP-WEBT archive; for questions regarding
  their availability, please contact the WEBT President Massimo Villata.}}
\author{A.~W.~Chen\inst{1,2,*}, F.~D'Ammando\inst{3,4},
  M.~Villata\inst{5}, C. M.~Raiteri\inst{5},
  M.~Tavani\inst{3,4}, V.~Vittorini\inst{2,3}, A.~Bulgarelli\inst{6},
  I.~Donnarumma\inst{3}, A.~Ferrari\inst{5,7}, A.~Giuliani\inst{1},  F.~Longo\inst{8},
  L.~Pacciani\inst{3},  G.~Pucella\inst{3},  S.~Vercellone\inst{1},
  A.~Argan\inst{3}, G.~Barbiellini\inst{8}, F.~Boffelli\inst{9},
  P.~Caraveo\inst{1}, D.~Carosati\inst{10}, P.~W.~Cattaneo\inst{9}, V.~Cocco\inst{3},
  E.~Costa\inst{3},
  E.~Del Monte\inst{3}, G.~De Paris\inst{3},
  G.~Di Cocco\inst{6},
  Y.~Evangelista\inst{3},
  M.~Feroci\inst{3}, M.~Fiorini\inst{1},
  T.~Froysland\inst{2,4}, M.~Frutti\inst{3},
  F.~Fuschino\inst{6}, M.~Galli\inst{11},
  F.~Gianotti\inst{6}, O.~M.~Kurtanidze\inst{12,13,14},
  C.~Labanti\inst{6}, I.~Lapshov\inst{3},
  V.~M.~Larionov\inst{15,16}, F.~Lazzarotto\inst{3},
  P.~Lipari\inst{17},
   M.~Marisaldi\inst{6}, M.~Mastropietro\inst{18}, S.~Mereghetti\inst{1}, E.~Morelli\inst{6},
  A.~Morselli\inst{19},
  M.~Pasanen\inst{20}, A.~Pellizzoni\inst{1}, F.~Perotti\inst{1},
  P.~Picozza\inst{19}, G.~Porrovecchio\inst{3}, M.~Prest\inst{21},
  M.~Rapisarda\inst{22}, A.~Rappoldi\inst{9}, A.~Rubini\inst{3},
  P.~Soffitta\inst{3}, M.~Trifoglio\inst{6},
  A.~Trois\inst{3}, E.~Vallazza\inst{8},
  A.~Zambra\inst{1},
  D.~Zanello\inst{17}, S.~Cutini\inst{23}, D.~Gasparrini\inst{23},
  C.~Pittori\inst{23}, P.~Santolamazza\inst{23}, F.~Verrecchia\inst{23},
  P.~Giommi\inst{23}, L.~A.~Antonelli\inst{23},
  S.~Colafrancesco\inst{23}, L.~Salotti\inst{24}
}

\institute{$^1$INAF/IASF--Milano, Via E.~Bassini 15, I-20133 Milano, Italy \\
$^2$CIFS--Torino, Viale Settimio Severo 63, I-10133 Torino, Italy \\
$^3$INAF/IASF--Roma, Via del Fosso del Cavaliere 100,
  I-00133 Roma, Italy \\
$^4$Dip. di Fisica, Univ. ``Tor Vergata'', Via della Ricerca
  Scientifica 1, I-00133 Roma, Italy \\
$^5$INAF, Osservatorio Astronomico di Torino, Italy\\
$^6$INAF/IASF--Bologna, Via Gobetti 101, I-40129 Bologna, Italy \\
$^7$Dip. di Fisica Generale dell'Universit\'a, Via Pietro Giuria 1, I-10125 Torino, Italy \\
$^8$Dip. di Fisica and INFN Trieste, Via Valerio 2, I-34127 Trieste, Italy\\
$^9$INFN--Pavia, Via Bassi 6, I-27100 Pavia, Italy\\
$^{10}$Armenzano Astronomical Observatory, Assisi, Italy \\
$^{11}$ENEA, Via Martiri di Monte Sole 4, I-40129 Bologna,
  Italy\\
$^{12}$Abastumani Astrophysical Observatory, Georgia\\
$^{13}$Astrophysikalisches Institut Potsdam, Germany\\
$^{14}$Landessternwarte Heidelberg-K\"onigstuhl, Germany\\
$^{15}$Astronomical Institute, St.-Petersburg State University, Russia\\
$^{16}$Pulkovo Observatory, Russia\\
$^{17}$INFN--Roma ``La Sapienza'', Piazzale A. Moro 2, I-00185 Roma,
  Italy\\
$^{18}$CNR, Istituto Metodologie Inorganiche e dei Plasmi, Area Ricerca
  Montelibretti (Roma), Italy\\
$^{19}$INFN--Roma ``Tor Vergata'', Via della Ricerca Scientifica 1,
  I-00133 Roma, Italy\\
$^{20}$Tuorla Observatory, University of Turku, Finland\\
$^{21}$Dip. di Fisica, Univ. dell'Insubria, Via Valleggio 11,
  I-22100 Como, Italy\\
$^{22}$ENEA--Roma, Via E. Fermi 45, I-00044 Frascati (Roma), Italy\\
$^{23}$ASI--ASDC, Via G. Galilei, I-00044 Frascati (Roma), Italy\\
$^{24}$ASI, Viale Liegi 26 , I-00198 Roma, Italy\\}

\offprints{A. Chen, \email{chen@iasf-milano.inaf.it} }

   \date{received: 10 July 2008; accepted: 18 August 2008}

% \abstract{}{}{}{}{}
% 5 {} token are mandatory

  \abstract
   {}
  % aims heading (mandatory
   {We report the $\gamma$-ray activity from the intermediate BL Lac S5 0716+714
     during observations acquired by the AGILE satellite in September and October 2007. These detections of activity were contemporaneous 
  with a period of intense optical activity, which was monitored
  by GASP--WEBT. This simultaneous optical
  and $\gamma$-ray coverage allows us to study in detail the light curves, time
  lags, $\gamma$-ray photon spectrum, and Spectral Energy Distributions (SEDs)
  during different states of activity.}
   {AGILE observed the source with its two co-aligned imagers,
the Gamma-Ray Imaging Detector (GRID) and the hard X-ray imager (Super-AGILE),
  which are sensitive to the
30~MeV--50~GeV and 18--60 keV energy ranges, respectively. Observations were
  completed in two different periods, the
  first between 2007 September 4 -- 23,
and the second between 2007 October 24 -- November 1.}
   {Over the period 2007 September 7 -- 12, AGILE detected $\gamma$-ray
  emission from the source at a significance level of 9.6-$\sigma$ with an average flux
  (E$>$100~MeV) of $(97 \pm 15) \times 10^{-8}$ photons cm$^{-2}$ s$^{-1}$,
  which increased by a factor of at least four within three days. No emission was
  detected by Super-AGILE for the energy range 18--60~keV to a 3-$\sigma$
  upper limit of 10 mCrab in 335 ksec. In October 2007, AGILE repointed toward
  S5 0716+714 following an intense optical flare, measuring an average flux of
 $(47 \pm 11) \times 10^{-8}$ photons cm$^{-2}$ s$^{-1}$ at a significance
  level of 6.0-$\sigma$. }
  % conclusions heading (optional)
{ The $\gamma$-ray flux of S5 0716+714 detected
  by AGILE is the highest ever detected for this blazar and one of the most
  intense $\gamma$-ray fluxes detected from a BL Lac object. The SED of
  mid-September appears to be consistent with the synchrotron
  self-Compton (SSC) emission model, but only by including two SSC components of
  different variabilities.}   
\authorrunning{A. Chen et al.}
\titlerunning{AGILE detection of variable $\gamma$-ray activity from S5 0716+714}
  \maketitle
%
%________________________________________________________________

 \section{Introduction} \label{introduction}

The source S5 0716+714 was discovered in 1979 as the optical
counterpart to an extragalactic radio source (K\"uhr
et al. 1981).  Two years later, it was classified as a BL Lac
(Biermann et al. 1981) because of
its featureless optical spectrum and high linear polarization. The optical
continuum was so featureless that every attempt to determine its spectroscopic redshift has
failed. However, by optical imaging of the underlying galaxy, Nilsson et
al. (2008) derived a redshift of z = 0.31 $\pm$ 0.08.

 According to its spectral energy distribution the source belongs to the intermediate BL Lac class, observations by BeppoSAX (Tagliaferri et
al. 2003, Giommi et al. 1999) and XMM-$Newton$ (Foschini et al. 2006, Ferrero et
al. 2007) provide evidence for a concave X-ray spectrum in the
0.1--10~keV band, which is a signature of
the presence of both the steep tail of the synchrotron emission and the flat
part of the Inverse Compton spectrum. The detection in the X-ray band of
fast variability only
in the soft X-ray component can be interpreted as a slowly
variable Compton component and a fast, erratic variable tail of the
synchrotron component.

In general, the variability of this blazar is strong in every band on both
long and short (intraday) timescales. The optical and radio
historical behaviour was analyzed by Raiteri et al. (2003), while the
EGRET telescope on the Compton Gamma-Ray Observatory (Hartman et
al. 1999) detected S5 0716+714 several times in the $\gamma$-rays (Lin et al. 1995; von Montigny
et al. 1995). The integrated flux above 100 MeV varied between (13 $\pm$ 5) and
(53 $\pm$ 13) x 10$^{-8}$ photons cm$^{-2}$ s$^{-1}$.

In this Letter, we present the analysis of the AGILE data
obtained during the S5 0716+714 observations in the period 2007 September --
October, in particular two flaring episodes: the first in mid-September, the
other on 2007 October 22-23. Preliminary results were communicated in Giuliani et al. (2007).

The intense
$\gamma$-ray flare detected by AGILE in mid-September triggered observations by the
GLAST-AGILE Support Program 
(GASP) of the WEBT\footnote[1]{\texttt{http://www.oato.inaf.it/blazars/webt/}\\ see e.g. Villata et al. (2006, 2007); Raiteri et al. (2006, 2007).}
(see Carosati et
  al., 2007). About one month later, the GASP observed a bright phase of the source,
triggering new AGILE and Swift observations. In the period from September to
October 2007, S5 0716+714 showed intense activity with strong optical
flaring episodes and a rare contemporaneous optical-radio outburst (Villata et
al. 2008).

The results of a multiwavelength campaign on S5 0716+714 with
simultaneous Swift and AGILE observations in October 2007 are discussed in
Giommi et al. (2008).
Throughout this paper, the quoted uncertainties are given at the
1--$\sigma$ significance level, unless otherwise stated.

%______________________________________________________________________________

\section{AGILE observation of S5 0716+714} \label{0716:pointing}

The AGILE scientific Instrument (Tavani et al. 2008)
is very compact and combines four
active detectors that provide broad-band coverage from hard X-rays
to $\gamma$-rays: a Silicon Tracker optimized for $\gamma$-ray imaging in the 30
MeV--50 GeV energy band (Prest et al. 2003); a
co-aligned coded-mask X-ray imager sensitive in the 18--60 keV energy band (Feroci et al. 2007); a
non-imaging Cesium Iodide Mini-Calorimeter sensitive in the 0.3--100 MeV
energy band (Labanti et al. 2006); and
a segmented Anticoincidence System (Perotti et al. 2006). 
%The combination of Silicon Tracker, Mini-Calorimeter and Anticoincidence
%System forms the Gamma-Ray Imaging Detector (GRID).

In September, during its Science
Performance Verification Phase the AGILE satellite devoted three weeks to the
observation of S5 0716+714 between 2007 September 4 14:58 UT and September 23
11:50 UT, for a total pointing duration of $\sim 16.9$
days$\footnote[2]{Between 2007  September 15 12:52 UT and September 16 12:42
  UT AGILE performed a calibration test on the Crab pulsar and S5 0716+714 was
  out of the Field of View of the satellite for two days.}$. In October, AGILE repointed toward the source and observed S5 0716+714 between
2007 October 24 9:47 UT and November 1 12:00 UT, for a total pointing
duration of $\sim$ 8.1 days.

%\begin{figure}
%\centering
%\includegraphics[angle=0,scale=0.29]{fig1.eps}
%    \caption[Map of S5~0716+714.]{Gaussian-smoothed counts map
%      of S5~0716+714
%      in Galactic coordinates integrated over the
%      observing period of most intense activity (2007 September 7 14:24 UT --
%      2007 September 12 12:00 UT) with the source at about 15$^{\circ}$ off-axis.
    %  The cross symbol is located at the \source{}
    %  radio coordinates.
%    \label{0716:fig:map}}
%\end{figure}
%______________________________________________________________________________

\section{Data Analysis and Results} \label{0716:dataanal}

Level--1 AGILE-GRID data were analyzed using the
AGILE Standard Analysis Pipeline. After the alignment of all data times
to Terrestrial Time (TT), an ad-hoc implementation of the Kalman Filter
technique was used to achieve track identification and event direction reconstruction. Subsequently, a
quality flag was assigned to each GRID event: G, P, S, or L, depending on
whether it was
recognized to be a $\gamma$-ray event, a charged particle, a single-track event,
or if its nature was uncertain, respectively. We selected only events flagged as
confirmed $\gamma$-ray events, while all events collected during the South
Atlantic Anomaly were rejected. We also rejected all $\gamma$-ray
events with reconstructed directions that formed angles with the satellite-Earth
vector smaller than 80$^{\circ}$; this reduced the $\gamma$-ray Earth
Albedo contamination by excluding regions within $\sim$ 10$^{\circ}$ of the
Earth limb. Counts, exposure, and Galactic background $\gamma$-ray maps were
created with a binsize of 0.$^{\circ}$3 $\times$ 0.$^{\circ}$3 for photons
of energies higher than 100 MeV.

The average $\gamma$-ray flux as well as the daily values were derived by the
maximum likelihood analysis of Mattox et al. (1993).
First, the entire period was analyzed to determine the diffuse
emission parameters, then, the source flux density was estimated
independently for each of the 1-day periods with the diffuse
parameters fixed at the values obtained in the first step.

%
%We ran the AGILE Maximum Likelihood procedure (ALIKE) on the whole
%observing period, in order to obtain a value for the average flux.
%
%

%\begin{figure}[!h] %%%%%%%%%%%%%%%%%%%%%%%%%%%%% FIG 2
%\centering
%  \includegraphics[angle=0,scale=0.45]{figcontour_2_200.eps}
%    \caption[Contours of S5~0716+714.]{Solid curve: AGILE 95\% maximum
%      likelihood contour level; dash-dot curve: AGILE 95\% error circle taking
%      into account both statistic and systematic uncertainties; dashed
%      curve: EGRET 95\% error ellipse (Mattox et al. 2001);
%      star: AGILE 95\% maximum likelihood contour level barycentre;
%      cross: radio position of S5~0716+714.
%    \label{0716:fig:contour}}
%\end{figure}

%_____________________________________________________________________________

%Figure~\ref{0716:fig:map} shows a Gaussian-smoothed counts map
%in Galactic coordinates integrated over the observing period of most intense activity, 7--12 September
%2007; the source is at about 15$^{\circ}$ off-axis and using the selections
%described above. 
The AGILE 95\% maximum likelihood contour level barycenter of the source is
$l=143.\!^{\circ}87$\,,\,$b=28.\!^{\circ}00$. The distance between this
position and the S5 0716+714 radio position ($l=143.\!^{\circ}98$\,,\,$b=28.\!^{\circ}02$) is $0.\!^{\circ}16$. 
The AGILE 95\% maximum likelihood contour level has a
semi-major axis $a=0.\!^{\circ}33$ and a
semi-minor axis $b=0.\!^{\circ}33$, whereas the overall AGILE error circle,
taking into account systematic errors, 
has a radius $r=0.\!^{\circ}58$.

\begin{figure}[!th] %%%%%%%%%%%%%%%%%%%%%%%%%%%%% FIG 2
    \centering
    \includegraphics[angle=0,scale=0.58]{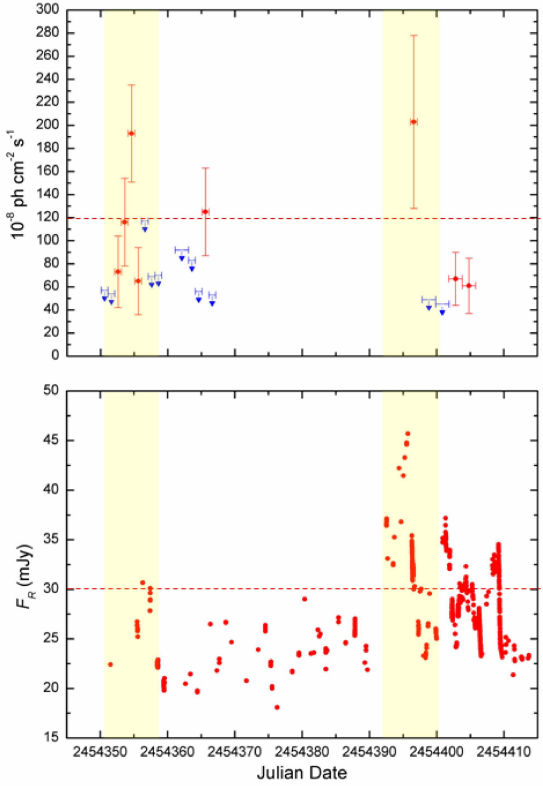}
    \caption[0716 lc]{% MJD 54340 = 7 Sep 2007
      In the top panel, the AGILE--GRID $\gamma$-ray light curve with
      1-day or 2-day resolution for fluxes in units of
      $10^{-8}$\,photons cm$^{-2}$ s$^{-1}$ for E $>$ 100 MeV. The downward
      arrows represent 2-$\sigma$ upper limits. In the bottom panel, the $R$-band optical light
      curve as observed by GASP-WEBT. In both panels, the mean flux level is
      highlighted with horizontal red dashed lines and the yellow shaded regions
      indicate the two high-activity periods in the $\gamma$-ray band.
    \label{0716:fig:lcagile}}
\end{figure}

The top panel in Fig.~\ref{0716:fig:lcagile} shows the $\gamma$-ray light curve
with 1 or 2 day resolution for photons of energy above 100~MeV. The downward arrows
represent 2-$\sigma$ upper limits. During the September observation, we note that S5 0716+714 is detected at high significance
on 1-day timescales, while other days have upper limits that are significantly
lower, clearly indicating strongly variable $\gamma$-ray activity.

The average $\gamma$-ray flux above 100 MeV during the period 2007 September
7--12 is $ F_{E > 100~\rm MeV}  = (97 \pm 15) \times 10^{-8}$\,photons cm$^{-2}$
s$^{-1}$ with a peak level of $ F_{E > 100~\rm MeV}  = (193 \pm 42) \times
10^{-8}$\,photons cm$^{-2}$ s$^{-1}$, and shows an increase in flux by a
factor of four in three days. The peak level $\gamma$-ray flux above 100 MeV,
observed in mid September, was the highest ever detected from this source.

Super-AGILE observed S5 0716+714 during 2007 September 7--12 for a total
on-source net exposure time of 335~ksec.
The source was not detected (above 5-$\sigma$) by the Super-AGILE
Iterative Removal Of Sources (IROS) algorithm, which was applied to the  image in
the 20--60~keV energy range. A 3-$\sigma$ upper limit of 10 mCrab was derived
from the observed count
rate by a study of the background fluctuations at the
position of the source and a simulation of the source and background
contributions with IROS.

A comparison between the $\gamma$-ray and optical light curves is shown in
Fig.~\ref{0716:fig:lcagile}; the bottom panel shows the $R$-band
optical light curve obtained by the GASP-WEBT. The results of the GASP-WEBT
multifrequency monitoring of 0716+714 in September--October 2007 are presented
in Villata et al. (2008)\footnote[3]{In addition to the data presented there, we show new GASP data for a couple of nights.}. 
%Unfortunately there is no data for the three days of the rise in $\gamma$-ray
%flux; however, the optical data show a fast rise within two days of the $\gamma$-ray peak.

The source in mid-October showed increasing optical flux, which reached a peak
of $F_R$ = 45.7 mJy on October 22.2 (Villata et al. 2008); at that time S5 0716+714,
even if rather off-axis ($\sim$ 50$^\circ$ from the axis) was seen by AGILE to
have a
high $\gamma$-ray flux. In particular, between
2007 October 22 12:33 UT and October 23
12:06 UT, the maximum likelihood analysis measured a flux of $ F_{E > 100 \rm MeV}  = (203 \pm 75) \times
10^{-8}$\,photons cm$^{-2}$ s$^{-1}$ at a significance level of 4.0-$\sigma$. We note, however, that AGILE has a high particle
background at high off-axis angles, and that the exposure time is relatively
short (only one day of observation).

After this flaring episode, AGILE observed the source with a dedicated
repointing at an off-axis angle of $\sim$ 15$^\circ$ between October 24 9:47 UT and November 1
12:00 UT. During this entire period, the AGILE-GRID detected a $\gamma$-ray flux above 100
MeV at a significance level of 6.0-$\sigma$ with a lower average
flux of
$ F_{E > 100 \rm MeV}  = (47 \pm 11) \times 10^{-8}$\, photons cm$^{-2}$
s$^{-1}$.

During the September--October observations, AGILE measured S5 0716+714 to have
two
different levels of activity. The $\gamma$-ray spectrum during the high activity state of mid-September
can be fitted with a power-law of photon index $\Gamma$ = 1.56
$\pm$ 0.30, while during the AGILE October ToO the source was in a low
$\gamma$-ray activity state and the photon index of the differential energy spectrum was $\Gamma$ = 1.95
$\pm$ 0.54 (see Fig.2). The photon index was obtained with the least squares
method by
considering only three energy bins: 100--200 MeV, 200--400 MeV and 400--1000 MeV.
\begin{figure}[!t]
\centering
\includegraphics[angle=0,scale=0.08]{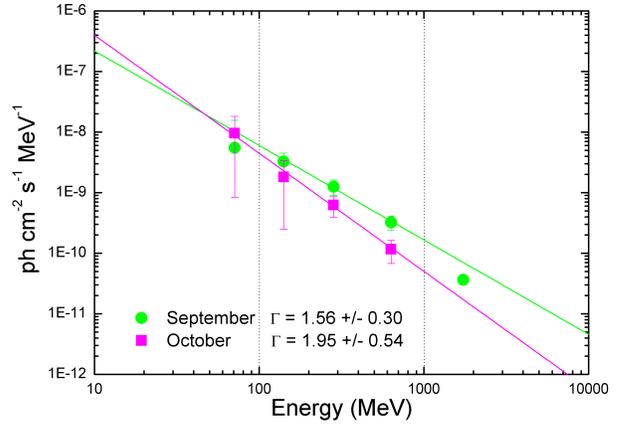}
    \caption[Gamma-ray spectrum of S5~0716+714.]{Gamma-ray photon spectrum
      of S5~0716+714 during the high state of mid-September 2007 (green line)
      and the low state of end October 2007 (magenta line).
    \label{0716:powerspectrum}}
\end{figure}

\section{Discussion} \label{0716:discussion}

To analyze the gamma-optical correlation, we have applied the Discrete Correlation Function
(DCF; see Edelson $\&$ Krolick (1988) and Hufnagel $\&$ Bregman (1992))
to the $\gamma$-ray and $R$-band light curves.
The DCF is a statistical method developed to analyze unevenly sampled data sets.
The $R$-band flux densities were averaged over 0.1 day bins to smooth the intranight variability.
The result is shown in Fig. 3. The DCF displays a significant peak (DCF $\sim$
0.9) for a time-lag of -1 day. Notwithstanding the large uncertainty due to poor $\gamma$-ray
sampling, this result suggests a possible delay in the $\gamma$-ray flux
variations with respect to optical variations of the order of 1 day. The uncertainty in the delay can be estimated by
Monte Carlo simulations based on the ``flux randomization / random subset
selection'' method (see Peterson et al. (2001) and Raiteri et al. (2003)). By
performing 2000 simulations we derived a 1-$\sigma$ uncertainty level in the lag of 1.1
days.

\begin{figure}[!t] %%%%%%%%%%%%%%%%%%%%%%%%%%%%% FIG 4
    \centering
    \includegraphics[angle=0,scale=0.40]{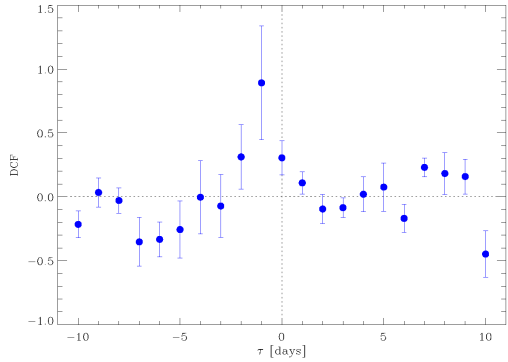}
    \caption[0716 DCFsep]{Discrete correlation function (DCF) between the
    $\gamma$-ray and $R$-band light curves for S5 0716+714 in September-October 2007.

    \label{0716:fig:dcs}}
\end{figure}

In Fig. 1, it is clear that most of the DCF signal originates from the quasi-simultaneity of the
$\gamma$-ray and optical peaks of late October (JD $\sim$ 2454396-397). As for
the September AGILE detection, the strong $\gamma$-ray flare lacks strictly
simultaneous optical observations since it occurred at both the start of the GASP operation and the optical observing season.

\begin{table}[!b]
\begin{center}
\caption{Parameters for the two SSC components.}
\begin{tabular}{|l|lll|}
\hline

                     &1$^{st}$ SSC comp &2$^{nd}$ SSC comp &Units \\
\hline
$\delta$             &14         &25     &       \\
$\Gamma$             &7.5       &15      &       \\
$R$                  &40        &40     & [$10^{15}$ cm]   \\
$B$                  &1         &0.5      &  [G]    \\
$\gamma_{\rm min}$   &200       &3 x 10$^{3}$     &     \\
$\gamma_{\rm break}$ &4 x 10$^{3}$  &6 x 10$^{3}$     &     \\
$p_{\rm  low}$        &2.0      &2.0    &     \\
$p_{\rm  high}$       &4.8      &4.8    &     \\
$n_{\rm e}$          &2.2         &0.8   & [cm$^{-3}$] \\
$\theta$             & 2         & 2          &  [deg] \\         
\hline
\end{tabular}
\end{center}
\end{table}

We note that when the $\gamma$-ray fluxes are $\la$ 120 $\times$
10$^{-8}$\, photons cm$^{-2}$ s$^{-1}$, the corresponding optical
flux densities are around 25--30 mJy. In contrast, the October
$\gamma$-ray peak reaching $\sim$ 200 $\times$ 10$^{-8}$\, photons
cm$^{-2}$ s$^{-1}$ has an optical counterpart of 40--45 mJy (see Fig. 1). This
suggests that a significant optical event occurred at the same time as the $\gamma$-ray
flare and in September was missed. Moreover, while the ratio between
the high and low $\gamma$-ray flux levels is about 2.5, in the
optical band the same ratio is of the order of 1.5. Hence, the gamma variability
appears to depend on the square of changes in optical flux density. This would favour a SSC interpretation, in which
the emission at the synchrotron and IC peaks is produced
by the same electron population, which self-scatters the
synchrotron photons. The 1-day time-lag in the high-frequency
peak emission found from the DCF could then be due to the light travel time of the
synchrotron seed photons that scatter the energetic electrons.

\begin{figure}[!h] %%%%%%%%%%%%%%%%%%%%%%%%%%%%% FIG 5
\centering
  \includegraphics[angle=0,scale=0.08]{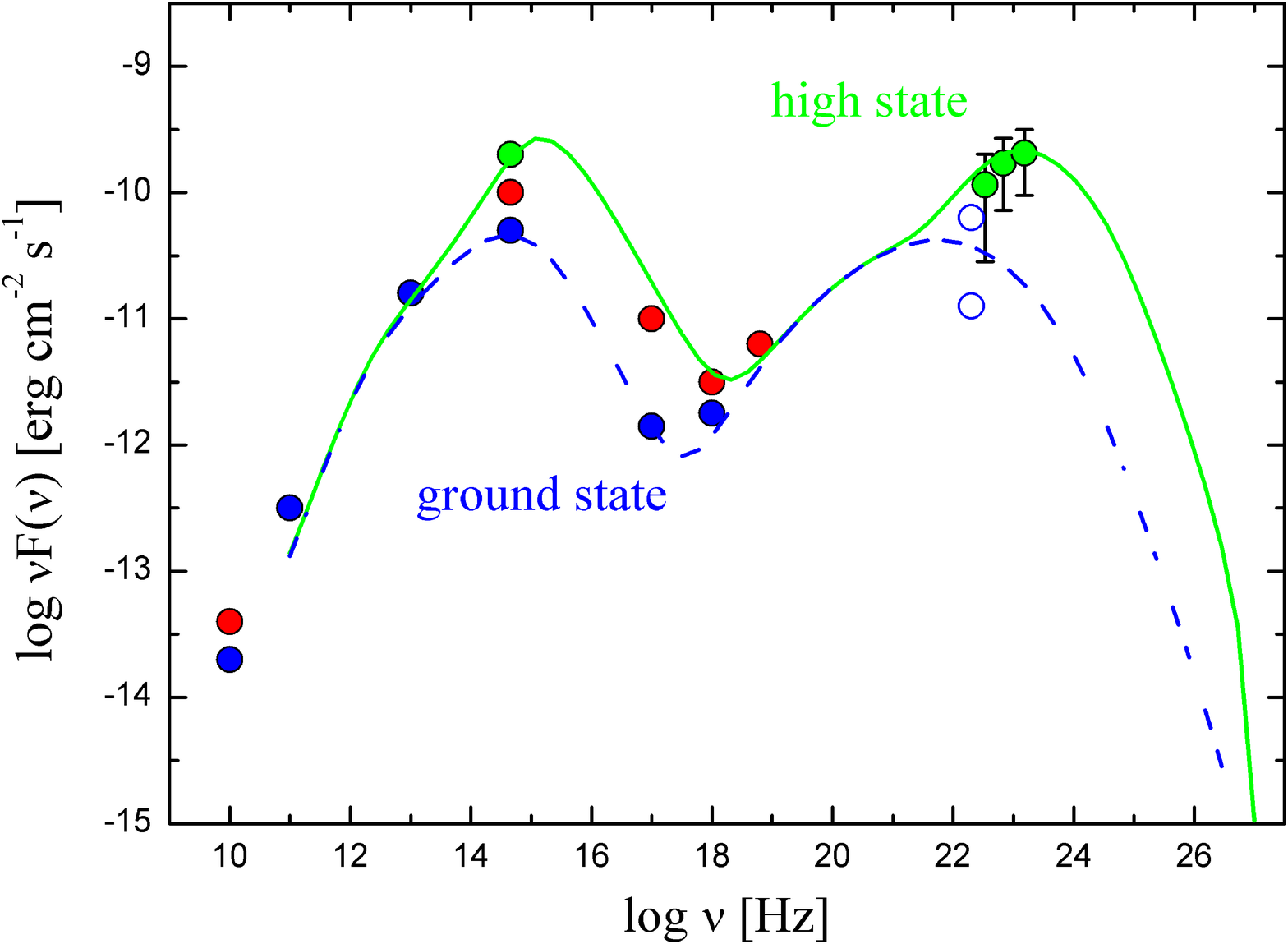}
     \caption[SED of S5~0716+714.]{
       The SED of S5 0716+714, including GASP-WEBT optical data
       quasi-simultaneous with a AGILE-GRID gamma-ray
       observation in September (green dots). Historical data over the entire
       electromagnetic spectrum relative to a ground state of the source and EGRET non-simultaneous data are
       represented with blue dots. Red dots represent historical data simultaneous with a high X-ray state.
     \label{0716:fig:spectrum}}
\end{figure}

The Spectral Energy Distribution for the AGILE and GASP-WEBT data
of September 2007 is shown in Fig.~\ref{0716:fig:spectrum} as green dots. The
blue dashed line shows a simple SSC model that fits simultaneous observations
of a ground state (see Tagliaferri et al. 2003
and references therein) and non-simultaneous EGRET data (empty blue circles).
Because the high state of mid-September 2007 cannot be fitted by a
one-zone SSC component alone, we used a model with two SSC components.

To the first SSC component that reproduces the ground state, we
add a second SSC component that dominates the optical and $\gamma$-ray
bands. Both the components are
reproduced  with a double power-law electron distribution: the spectral index
is $p_{low}$ from $\gamma_{min}$ to $\gamma_{break}$
and $p_{high}$ above $\gamma_{break}$. The parameters of the two SSC
components are reported in detail in Table 1.

We cannot exclude a second component due to an external seed
photon field, for example mirrored by a putative broad line region, which could also account for the possible
1 day time lag. Nevertheless, the large
amplitude of $\gamma$-ray variability with respect to that of the optical favors a SSC explanation.

%Simultaneous Swift and AGILE observations in October are discussed in Giommi
%et al. (2008).

The luminosities observed in the optical and $\gamma$-ray ranges are both $10^{48}$ erg s$^{-1}$, with a dissipated power in the jet
rest-frame of $2 \times 10^{43} (\delta/15)^{-4}$ erg s$^{-1}$. This output is
significantly large with respect to other BL Lacs, and we estimate a global power
transported into the jet $L_{tot}>3 \times 10^{45}$ erg s$^{-1}$. This
may exceed the maximum power generated by a
spinning black hole of mass $10^{9}$ M$_\odot$ in most widely known
models (see e.g. Cavaliere $\&$ D'Elia 2002).

\begin{acknowledgements}
The AGILE Mission is funded by the Italian Space Agency (ASI) with
scientific and programmatic participation by the Italian Institute
of Astrophysics (INAF) and the Italian Institute of Nuclear
Physics (INFN). We wish to express our gratitude to
the Carlo Gavazzi Space, Thales Alenia Space, Telespazio and ASDC/Dataspazio
Teams that implemented the necessary procedures to carry out the
AGILE re-pointing. 
%The optical data presented in this paper are
%stored in the GASP-WEBT archive; for questions regarding their availability,
%please contact the WEBT President Massimo Villata.
{\it Facilities:AGILE}
\end{acknowledgements}

\Online
%\begin{appendix}
\begin{figure*}
\centering
\includegraphics[angle=0,scale=0.4]{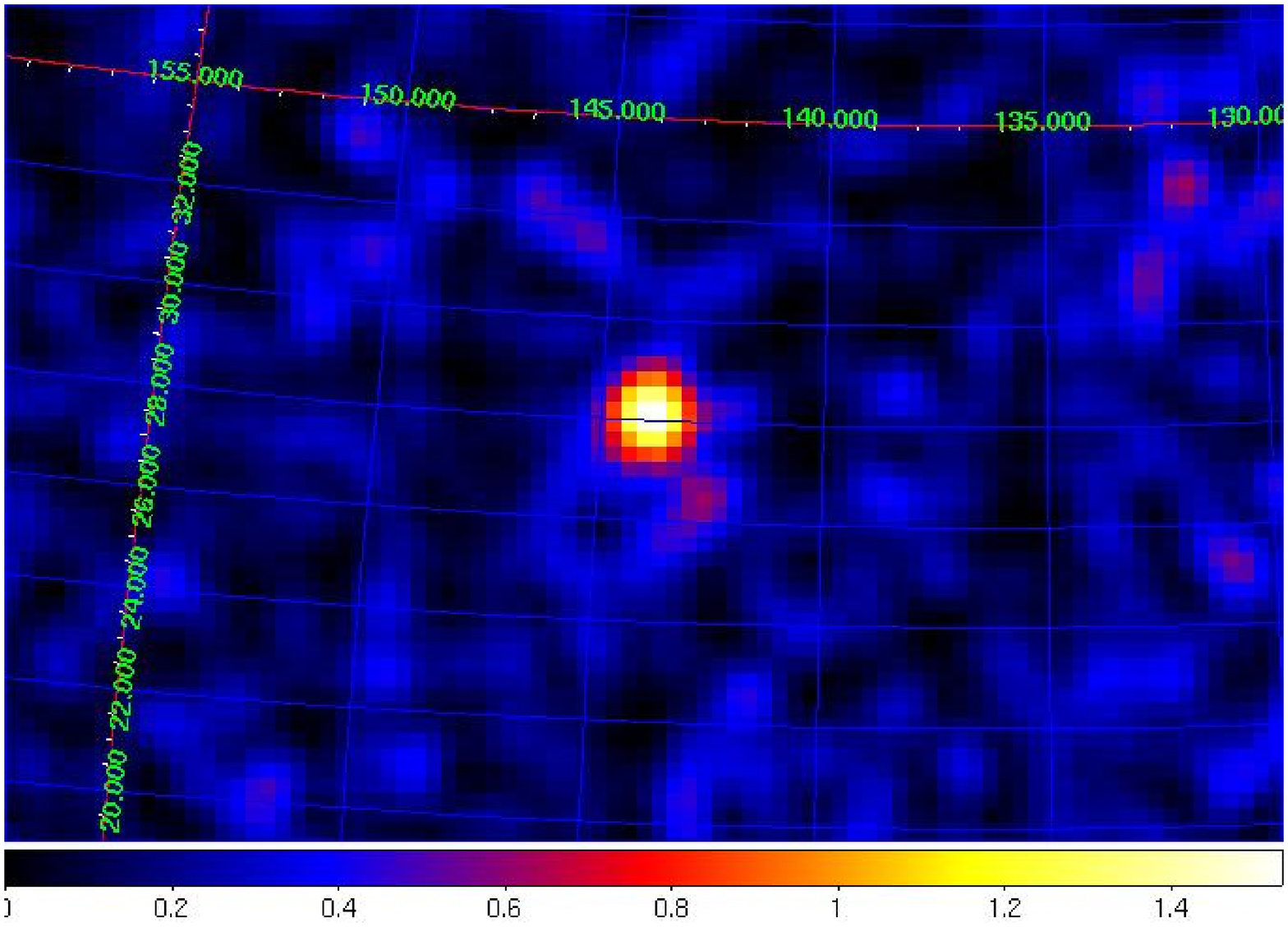}
    \caption{Gaussian-smoothed counts map
      of S5~0716+714
      in Galactic coordinates integrated over the
      observing period of most intense activity (2007 September 7 14:24 UT --
      2007 September 12 12:00 UT) with the source at about 15$^{\circ}$ off-axis.}
\end{figure*}

\end{document}